\documentclass[12pt]{iopart}
\usepackage{epsfig, epsf}

\newcommand{\beqs}{\begin{eqnarray*}}

\newcommand{\eeqs}{\end{eqnarray*}}

\newcommand{\beq}{\begin{equation}}
\newcommand{\eeq}{\end{equation}}

\maketitle
\begin{document}

\title{Correlated Percolation}
\author{Antonio Coniglio$^{a,b}$ and Annalisa Fierro$^{b}$}

\address{${}^a$
Dipartimento di Fisica ``Ettore Pancini'', Universit\`a di
Napoli ``Federico II'', Complesso Universitario di Monte
Sant'Angelo, via Cintia, Naples, Italy}
\address{${}^b$ CNR-SPIN, Naples, Italy}
\section*{\bf Article Outline}
\begin{itemize}
\item[]
\begin{itemize}
\item[1.] Definition of the Subject and Its Importance
\item[2.] Introduction
\item[3.] Random Percolation
\begin{itemize}
\item[3.1.] Scaling and hyperscaling
\item[3.2.] Breakdown of hyperscaling
\item[3.3.] Cluster structure
\item[3.4.] Surfaces and interfaces
\end{itemize}
\item[4.] Percolation in the Ising Model
\begin{itemize}
\item[4.1.] Ising Clusters
\item[4.2.] Ising Droplets
\item[4.3.] Droplets in $2$ and $3$ dimensions
\item[4.4.] Droplets in an external field
\item[4.5.] Exact relations between connectivity and thermal properties
\item[4.6.] Ising Droplets above $d=4$
\item[4.7.] Generalization to the $q$-state Potts Model
\item[4.8.] Fractal structure in the Potts Model: Links and blobs
\item[4.9.] Fortuin Kasteleyn-Random Cluster Model
\end{itemize}
\item[5.] Hill's clusters
\item[6.] Clusters in weak and strong gels
\item[7.] Scaling behaviour of the viscosity
\item[8.] Future directions
\end{itemize}
\item[]  Appendix
\begin{itemize}
\item[A.] Random Cluster Model and Ising droplets
\begin{itemize}
\item[A.1.] Random Cluster Model
\item[A.2.] Connection between the Ising droplets and the Random Cluster Model
\end{itemize}

\end{itemize}
\item[] Bibliography
\end{itemize}

\section{\bf Definition of the Subject and Its Importance}
Cluster concepts have been extremely useful in elucidating many problems in
physics. Percolation theory provides a generic framework to study the behavior
of the cluster distribution. In most cases the theory predicts a geometrical
transition at the percolation threshold, characterized in the percolative phase
by the presence of a spanning cluster, which becomes infinite in the
thermodynamic limit. Standard percolation usually deals with the problem when
the constitutive elements of the clusters are randomly distributed. However
correlations cannot always be neglected. In this case correlated percolation is
the appropriate theory to study such systems. The origin of correlated
percolation could be dated back to 1937 when Mayer \cite{Mayer}
proposed a theory to describe the condensation from a gas to a liquid in terms
of mathematical clusters (for a review of cluster theory in simple fluids see
\cite{sator}). The location for the divergence of the size of these
clusters was interpreted as the condensation transition from a gas to a liquid.
One of the major drawback of the theory was that the cluster number for some
values of thermodynamical parameters could become negative. As a consequence
the clusters did not have any physical interpretation \cite{fisher}.
This theory was followed by Frenkel's phenomenological model \cite{frenkel},
in which the fluid was considered as made of non interacting physical
clusters with a given free energy. This model was later improved by Fisher
\cite{fisher}, who proposed a different free energy for the clusters,
now called droplets, and consequently a different scaling form for the droplet
size distribution. This distribution, which depends on two geometrical
parameters, $\sigma$ and $\tau$, has the nice feature that the mean droplet size
exhibits a divergence at the liquid-gas critical point. Interestingly the
critical exponents of the liquid gas critical point can be expressed in terms of
the two parameters, $\sigma$ and $\tau$, and are found to satisfy the standard
scaling relations proposed at that time in the theory of critical phenomena.


\section{\bf Introduction}
Fisher's droplet model was very successful, to describe the
behavior of a fluid or of a ferromagnet near the critical point,
in terms of geometrical cluster. However the microscopic
definition of such a cluster, in a fluid or ferromagnet was still
a challenge. While the exact definition in a continuum fluid model
is still an open problem, a proper definition in the Ising model
or lattice gas model has been provided. A first attempt to define
a cluster in the Ising model which had the same properties of
Fisher's droplet model was to consider a cluster as set of
parallel spins. In two dimensions in fact these clusters seemed to
have the properties of Fisher's droplets, i.e. the mean cluster
size of these clusters were found to diverge at the Ising critical
point on the basis of numerical analysis \cite{binder}. This
result was later proved rigorously \cite{15.,17.}. However the
critical exponents for the mean cluster size in $2d$ was found to
be larger than the corresponding critical exponent of the
susceptibility \cite{sykes}, contrary to the requirement of
Fisher's droplet model. Moreover numerical simulations in $3d$ and
analytical result on the Bethe lattice showed that the critical
point and the percolation point of such clusters were different. It
was clear then that the clusters made of nearest neighbors 
parallel spins were too big to describe correlated regions. It was
then proposed \cite{CK} a different definition of clusters
obtained by breaking the clusters of parallel spins by introducing
fictitious bonds with a probability $p_b$ between parallel spins.
The new clusters are defined as a maximal set of parallel spins
connected by bonds. For a particular choice of $p_b \equiv p =
1-e^{-2J/k_BT}$ it was shown that these clusters (Coniglio-Klein
droplets) have the same properties of Fisher's droplets, namely
their size diverges at the Ising critical point with Ising
exponents. Note that the bonds are only fictitious and do not
change the energy of the spins. They only have the role of
breaking the clusters made of parallel spins. Some years earlier
Kasteleyn and Fortuin defined a random cluster model, obtained
starting from an Ising model and by changing the spin interaction
$J$ in $J= \infty$, with probability $p$, and $J$ into $J=0$, with
probability $1-p$. They showed that the partition function of this
modified model, called random cluster model, coincides with the
partition function of the original Ising model. In the random
cluster model the clusters are defined as maximal set of spins
connected by infinite interactions. Although these clusters have
the same properties of the droplet model, they were defined in the
random cluster model, and for this reason these clusters were not
associated to the droplets of the Ising model. It was only after
Swendsen and Wang \cite{SW87} introduced a cluster dynamics based
on the Kasteleyn and Fortuin formalism, that it was formally shown
\cite{conigliodiliberto} that the distribution of the
Coniglio-Klein (CK) droplets are the same as the distribution of
Kasteleyn-Fortuin (KF) clusters in the random cluster model. For
this reason often the CK droplets and the KF clusters are
identified, however the different meaning should be kept in mind.

A further development was obtained when the fractal structure of the
droplets was studied not only for the Ising model but for the full
hierarchy of the $q$-state Potts model, which in the limit $q=1$ gives the
random percolation problem. It was shown that the critical
droplets of the Potts model have the same structure, made of links
and blobs, as found for the clusters in the random percolation
problem. One of the consequences of this study was a better
understanding of scaling and universality in terms of geometrical
cluster and fractal dimension \cite{BM}.

The cluster approach to the phase transitions lead also to a deeper
understanding of why critical exponents do not depend on
dimensionality above the upper critical dimension, and coincide with
mean field exponents. It was in fact suggested \cite{coniglio85} that,
at least for random percolation, the 
mean field behavior is due to the presence of an infinite multiplicity of
critical clusters at the percolation point. This suggests that similar results 
may be also extended to thermal problems.

Although the original interest in the field of correlated
percolation was the study of critical phenomena in terms of
geometrical concepts, later it was suggested that correlated
percolation could be applied to the sol-gel transition, in
particular when correlation was too large to be neglected. In many
cases in fact the sol-gel transition, which is based on long range
connectivity and percolation transition, interferes with large
density fluctuation or critical point. The interplay between
percolation points and critical points gives rise to interesting
phenomena which are well understood within the concepts of
correlated percolation. Correlated percolation has been studied also
in systems with different types of long range correlation
\cite{weinrib}, and has been applied to many other fields such as nuclear
physics \cite{nuclear}, Gauge Theory \cite{gauge} and O(n)
models \cite{On}, fragmentation \cite{fragm}, urban
growth \cite{hernan}, random resistor
network \cite{bastiansen}, interacting colloids \cite{int_col},
biological models \cite{abete}.

In Sect. \ref{RP} we introduce random percolation concepts.
In Sect. \ref{CP} in the context of the Ising model it is shown 
how clusters have to be defined in order to describe correlated regions
corresponding to spin fluctuations. In Sect.s \ref{CLGIM} and \ref{droplets}
the Ising clusters and droplets are respectively introduced, and in Sect. 
\ref{IDAD} it is shown how the mapping between thermal properties and 
connectivity breaks down below $T_c$ above $d=4$. 
In Sect. \ref{potts} the results found for the Ising model are extended
to the $q-$state Potts model, and in Sect. \ref{links_blobs} the fractal
structure is studied in terms of links and blobs. In Sect. \ref{FKRCM}
Fortuin Kasteleyn-Random Cluster Model is presented, and the connection with
the Coniglio-Klein droplets is further developed in Appendix A.
In Sect. \ref{secthill} the possibility to extend the definition of droplets to
simple fluids is discussed. In Sect. \ref{gel} the mechanism, leading to the
formation of bound states in gelling systems, is considered, and in Sect.
\ref{sect_viscosity} the effect that finite bond
lifetime has on the behaviour of viscosity in weak or colloidal gels.
Finally future directions and open problems are discussed
in Sect. \ref{future}.
\section{\bf Random Percolation}
\label{RP}
In this section we define some connectivity quantities and present some results
in the context of random percolation, which we will use in the following 
sections, where the correlated
percolation will be presented.

Consider a $d$-dimensional hypercubic lattice of linear dimension
$L$. Suppose that each edge has a probability $p$ of being occupied
by a bond. For small values of $p$, small clusters made of sites
connected by nearest-neighbour bonds are formed. Each cluster is
characterized by its size or mass $s$, the number of sites in the
cluster. For large values of $p$ in addition to small clusters we
expect a macroscopic cluster that connects the opposite boundaries.
This spanning cluster becomes infinite as the system size becomes
infinite. For an infinite system there exists a percolation
threshold $p_c$ below which only finite clusters are present.

In order to describe the percolation transition \cite{Essam,SA,bunde}, one defines:
an order parameter, $P_\infty (p)$, as the density of sites in
the infinite cluster, the mean cluster size, $S(p)$, of the finite clusters,
and the average number of clusters, $K(p)$.

These quantities can be related to the average number of clusters of $s$ sites
per site, $n (s,p)$, and near the percolation threshold the critical
behaviour is characterized by critical exponents:
\begin{equation}
K(p)|_{sing}  =  \sum n (s,p)|_{sing} \sim |p-p_c|^{2-\alpha_p}, \\
\end{equation}
\begin{equation}
P_{\infty}(p)  =  1 - \sum sn (s,p) \sim \left\{ \begin{array}{ll}
                               0                    & \mbox{if $p<p_{c}$}\\
                               (p-p_{c})^{\beta_p}  & \mbox{if $p>p_{c}$},\\
                             \end{array}
                     \right.
\end{equation}
\begin{equation}
S (p)  =  \sum s^2 n(s,p) \sim |p-p_c|^{-\gamma_p},
\end{equation}
where the sum is over all finite clusters, and in Eq. (1) only
the singular part has been considered.
Finally one can define the pair connectedness function $p^f_{ij}$ as the 
probability that $i$ and $j$ are in the same finite cluster through
\beq
\xi^2 (p)=\frac{\sum r^2_{ij} p^f_{ij}}{\sum p^f_{ij}}.
\eeq
The connectedness length, $\xi (p)$, which is the critical radius of the 
finite clusters, diverges
as
\beq
\xi  \sim |p-p_c|^{-\nu_p}~.
\label{xi}
\eeq
The critical exponents defined in Eq.s (1)-(4) are not all
independent. Scaling relations can be derived among them as
for ordinary second phase transitions. These scaling laws
are intimately related to the property of the incipient
infinite cluster of being a self similar fractal \cite{BM} to
all length scales. The mass, $s^*$, of a typical cluster of
linear dimension, $\xi$, scales as $s^* \sim \xi^{D_p}$, where
$D_p$ is the fractal dimension of the cluster.

\subsection{\bf Scaling and hyperscaling}
\label{SH}
To obtain scaling laws, following Kadanoff's original idea,
we perform \cite{coniglio85} the following three steps: (i) divide the system
into cells of linear dimension $b$, (ii) coarse grain by some
suitable rule, (iii) rescale the lengths by a factor $b$. The
result is renormalized system where the size of the large
clusters $s$ has been reduced by factor $b^{D_p}$ and all
lengths by a factor $b$:
\beq
L' = L/b\, , ~~~~~\xi =\xi' / b\, , ~~~~~ s' =s/b^{D_p}~.
\label{elleprimo}
\eeq
Assuming that the large clusters do not interpenetrate, the
sum over the large clusters in an interval between ($s,
s+\Delta s$) must be the same before and after rescaling, i.e
\beq
N (s, \xi) \Delta s \sim N (s', \xi') \Delta s'
\label{ennesse}
\eeq
where
$ N (s, \xi) / L^d =\overline{n} (s, \xi)$ is the number of clusters of
$s$ sites per unit volume. Dividing by the volume $L^d$,
from (\ref{elleprimo}) we obtain
\beq
\overline{n} (s, \xi) =b^{-d -D_p} \overline{n} (sb^{-D_p}, \xi b^{-1})~.
\label{ennesse2}
\eeq

Choosing $b =s^{1/D_p}$ from (\ref{ennesse2}) we obtain
\beq
n(s,p) =s^{-\tau_p} f(( p - p_c) s^{\sigma_p})
\label{ns}
\eeq
where $n(s,p) = \overline{n} (s, \xi)$ and
\beq
\tau_p = \frac{d}{D_p} +1~~~~~~~~ \sigma_p =\frac{1}{\nu_p D_p}~.
\label{doppia}
\eeq
Eq. (\ref{ns}) exhibits the scaling form postulated by
Stauffer \cite{SA,bunde}. From (1), (2) and (\ref{doppia}) we have:
\beq
2-\alpha_p =\frac{\tau_p - 1}{\sigma_p},~~~~~ \beta_p = \frac{\tau_p -2}{\sigma_p},~~~~~ -\gamma_p = \frac{\tau_p -3}{\sigma_p}
\label{menalpha}
\eeq
and
\beq
\tau_p=2+\frac{\beta_p}{\beta_p+\gamma_p},~~~~~
\sigma_p=\frac{1}{\beta_p+\gamma_p},
\eeq
from which the following scaling relation are obtained:
\beq
\alpha_p + 2 \beta_p + \gamma_p =2,
\label{scaling1}
\eeq
\beq
\frac{1}{\nu_p} (\beta_p +\gamma_p) =D_p.
\label{unsunu}
\eeq

From (\ref{doppia}), (\ref{menalpha}) one can also
find relations which contain the Euclidean dimensionality
$d$ called hyperscaling relation:
\beq
2 - \alpha_p =\nu_p d~,
\label{2menalpha}
\eeq
\beq
d- \frac{\beta_p}{\nu_p} =D_p ~.
\label{d-b}
\eeq
Eq. (\ref{d-b}) was originally suggested in Ref. \cite{kirk}.
In $2d$ exact results give $\tau_p=187/91$ and $\sigma_p=36/91$,
and in $3d$ the best estimates $\tau_p \simeq 2.18$,
$\sigma_p\simeq 0.45$.
From mean field theory \cite{harris} we know that for any $d$ above the upper
critical dimension $d_c =6$, the critical exponents coincide with the mean field
ones, namely $-\alpha_p = \beta_p =
\gamma_p =1$, $\nu_p =\sigma_p =1/2$ and $\tau_p =5/2$. These
exponents satisfy the scaling relation (\ref{scaling1}),
but fail to satisfy the hyperscaling relation
(\ref{2menalpha}) except for $d =6$.

Moreover, while Eq. (\ref{unsunu}), for any $d>6$, shows that the
fractal dimension is stacked at the value $D_p=4$, the
hyperscaling relation (\ref{d-b}) breaks down for $d>6$.

\subsection{\bf Breakdown of hyperscaling}
\label{BH}
By following a less conventional scaling approach, here we want to
propose a geometrical interpretation of
hyperscaling, why it breaks down above $d_c$, and why the hyperscaling breakdown
occurs when mean field becomes valid \cite{coniglio85}.

Let us assume that the singular behaviour
comes only from the critical clusters. Say
$N_\xi$ is the number of such clusters in a volume of
the order $\xi^d$. The singular part of the cluster number is given by
\beq
\frac{N_\xi}{\xi^d} \sim \xi^{\frac{\alpha_p - 2}{\nu_p}}~.
\label{kappap}
\eeq
At the same time, the density of sites in the infinite
cluster $P_\infty \sim |p-p_c|^{\beta_p}$ scales as the total mass
of the spanning clusters $N_\xi s^*$ in a volume of linear
dimension $\xi$ divided by the volume $\xi^d$, namely
\beq
\frac{N_\xi  \xi^{D_p}}{\xi^d} \sim \xi^{-\beta_p/\nu_p},
\label{ennexi}
\eeq
where we have used $s^* \sim \xi^{D_p}$. Similarly the mean cluster size:
\beq
\frac{N_\xi  \xi^{2D_p}}{\xi^d} \sim \xi^{\gamma_p/\nu_p}.
\label{ennexi2}
\eeq

These equations lead to the scaling relations, Eq.s
(\ref{scaling1}) and (\ref{unsunu}). Now if $N_\xi $ is of the
order of unity, we recover the hyperscaling relations
\cite{essam}, Eq.s (\ref{2menalpha}) and (\ref{d-b}), while if
$N_\xi$ diverges hyperscaling breaks down.  We know that for
dimension $d$ above $d_c=6$, $n(s,p) \sim s^{-5/2} e^{-(p-p_c)^2
s}$ for large $s$. Therefore $N_\xi =\xi^d \sum n (s,p) \sim
\xi^{d-6}$, where $\xi \sim |p -p_c|^{-1/2}$. This calculation
shows that, above $d_c$, $N_\xi$ diverges and hyperscaling breaks
down, and from Eq.s (\ref{kappap}) and (\ref{ennexi}) the
hyperscaling relations are replaced by $2-\alpha_p =6~\nu_p$ and
$D_p=6-\beta_p/\nu_p$, which in fact are satisfied for mean field
exponents.

The more standard scaling approach of previous section must be modified
taking into account that for $d>6$ the large number of clusters will be
reduced by a factor $b^{6-d}$, then Eq. (\ref{ennesse}) will
be modified as $N (s,\xi)\Delta s =b^{6-d} N(s', \xi') \Delta s'$
which still leads to all the Eqs. (\ref{ennesse}) - (\ref{unsunu}),
except that $d$ is replaced everywhere by 6.  In particular, both
Eqs. (\ref{unsunu}) and (\ref{d-b}) give a fractal dimension $D_p=4$.

The multiplicity of infinite clusters above $d_c$ was numerically
shown in Ref.s \cite{lda,fortunato}. The average
(finite) number $N_\xi$ of distinct clusters below $d_c$
have been estimated theoretically and calculated numerically
\cite{aizenman,hu1,stauffer2}.

Consider now a critical cluster for $d>d_c$
just below $p_c$ and its center of mass, 0.
Say $\xi_1$ the distance from 0, below which the cluster
has not been penetrated by the other critical clusters.
This length can be obtained by equating the mass density
inside the region of radius $\xi _1$ to the mass density inside the
region of radius $\xi$, $N_{\xi}\xi^{4-d} = \xi_1^{4-d}$,
which gives $\xi_1 \sim \xi^{2/(d-4)}$.

If $\rho (r)$ is the density profile defined
as the mass density of all the critical clusters at a distance $r$ from 0,
we expect that
the density profile behaves as a
power law $r^{d-4}$ for $r < \xi_1 $, as it should be for an object with fractal
dimension $D_p=4$ and as a constant for  $r > \xi_1 $  due to the
penetration of the other critical clusters.
Consequently we can make the following scaling
Ansatz \cite{aharony}:
\beq
\rho (r) =\frac{1}{r^{d-4}} \, f (r/\xi_1)\, ,
\label{rhor}
\eeq
where $f (x) \sim \mbox{const}$ for $x <1$ and $\sim x^{d-4}$
for $x >1$.

In conclusion, while for $d<6$ the density of the order parameter
fluctuates over a distance of the order $\xi$, for $d> 6$, where
mean field holds, the fluctuations are damped by the presence of
infinitely many interpenetrating clusters, and the density of the
order parameter crosses over from a power law (fractal) regime to
an homogeneous regime at a distance $\xi_1 \ll \xi$.

The mean field solution is therefore a consequence of the
presence of infinitely many interpenetrating clusters which
suppress the spatial fluctuation of the order parameter. The
condition for the validity of mean field theory is then given by
$N_\xi \gg 1~.$
Using Eq.s (\ref{ennexi}) and (\ref{unsunu}) this condition
implies
\beq
N_\xi^{-1} \sim \frac{\xi^{\gamma_p / \nu_p}}{\xi^d \xi^{-\beta_p /
\nu_p}} \sim \frac{\langle \Delta M^2 \rangle}{\xi^d M^2} \ll 1 \, ,
\label{implies}
\eeq
where $M$ and $ \langle \Delta M^2\rangle$ are the order
parameter and the fluctuations of the order parameter (here we used that
the mean cluster size $S(p)$ has the same critical behaviour as the
fluctuations of the order parameter \cite{con_stau,chayes1}).
Interestingly enough Eq. (\ref{implies})
coincides with Ginzburg criterion for the validity of mean field theory.

\subsection{\bf Cluster structure}
\label{CS}

{\em Nodes and Links}.~~~ In the previous Sections we have shown
that the Incipient Infinite Cluster (IIC) is a fractal. Here we
want to show in more details the internal structure of the IIC.
A very useful nodes and links picture for the infinite cluster just above $p_c$
was introduced by Skal and Shklowskii \cite{skal} and de
Gennes \cite{DG}.
In this picture the infinite cluster
consists of a superlattice made of nodes, separated by a distance
of the order of $\xi$, connected by macrobonds.  Just below
$p_c$ the structure of the very large cluster, the IIC, was
expected to have the same structure as the macrobonds.

Later on, in 1977, Stanley \cite{stanley} made
the important observation that in general for each configuration
of bonds at $p_c$ the IIC can be partitioned in three categories.
By associating an electric unit resistance to each
bond, and applying a voltage between the ends of the cluster, one
distinguishes the dangling bonds which do not carry current
(yellow bonds). The remaining bonds are the backbone bonds. The
backbone can be partitioned in singly connected bonds (red bonds)
and all the others, the multiply connected bonds, which lump
together in ``blobs" (blue bonds). The red bonds, which carry the
whole current, have also the property that if one is cut the
cluster breaks in two parts. This partition in three types of
bonds is very general and can be done for any cluster or
aggregate.

The next major problem was to determine whether the blobs are or
not relevant. In the nodes and links picture the assumption is
that the blobs are irrelevant and only links are relevant.
A further elaboration \cite{stanley} assumed
that the backbone close to $p_c$ would reduce to a
self avoiding walk chain, which implies that the blobs are not relevant.
This self avoiding walk Ansatz received a large amount
of attention, since it predicted a value for the crossover
exponent of the dilute Heisenberg ferromagnetic model near the
percolation threshold in $2d$, in good agreement with the
experimental data, although the prediction for the dilute Ising
crossover exponent did not agree as well with the data \cite{BC}.

\noindent {\em Syerpinsky gasket: a model without links}.~~~In 1981 a
completely alternative model was proposed by Gefen {\em et al.} \cite{gefen}.
Based on the observation that in a computer simulation the red
bonds were hardly seen, they proposed an alternative model, the
Syerpinsky gasket, that represents the opposite extreme of the
nodes and links picture. It has a self-similar structure but only
multiply connected bonds are present. A great advantage of this
model is that it can be solved exactly. It also gives good
prediction for the fractal dimension of the backbone, but it
fails to predict the correct value for the dilute Ising crossover
exponent \cite{BC}.

\noindent {\em Nodes, links and blobs}.~~ Motivated by all these
conflicting models, some rigorous results were presented which led
unambiguously to the nodes links and blobs picture of the infinite
cluster \cite{coniglio81,coniglio82}, in which both links and blobs are relevant
below $d=6$, while only links are relevant above $d=6$ or in mean field. 
In particular the following
relation was proven for any $p$ and for any lattice in any
dimension: \beq p\, \frac{dp_{ij}}{dp} \, =  \lambda_{ij}
\label{pfrac} \eeq where $p_{ij}$ is the probability that $i$ and
$j$ are connected, $\lambda_{ij}$ is the average number of red bonds
between $i$ and $j$, such that if one is cut, $i$ and $j$ would have
been disconnected. From Eq. (\ref{pfrac}) it is possible to
calculate the average number $L_{ij}$ of red bonds between $i$ and
$j$ under the condition that $i$ and $j$ are in the same cluster:
\beq L_{ij} = \lambda_{ij}  \, / \, p_{ij}~. \label{ij} \eeq From
the scaling form of $p_{ij} = r_{ij}^{-d+2-\eta_p} \, f (r_{ij} /
\xi)$ it follows \beq L_{ij} = r_{ij}^{1/\nu_p} \, f_1 (r_{ij}
/\xi)\, , \label{scaling} \eeq where $f_1 (x)$ is related via Eq.s
(\ref{pfrac}) and (\ref{ij}) to $f(x)$ and goes to a constant for $x
\ll 1$. In particular, by putting $r_{ij} =\xi$ in Eq.
(\ref{scaling}) we obtain \beq L_R \sim \xi^{1/ \nu_p}\, ,
\label{sim} \eeq where $L_R \equiv L (r_{ij} =\xi)$ is the average number of
red bonds between two points separated by a distance of the order of
$\xi$. From Eq. (\ref{sim}) it follows that the fractal dimension of
the red bonds is $D_R = 1/\nu_p$. An immediate consequence is that
not only the red bonds are relevant but also the number of bonds
$L_B$ in the blobs diverge. For more details, see
\cite{coniglio81,coniglio82}. Later Eq. (\ref{sim}) was confirmed
numerically by Pike and Stanley \cite{PS} in $d=2$. Although the
links are much less in number than the backbone bonds, they can be
detected experimentally, in fact it can be shown
\cite{coniglio81,coniglio83} that only the links determine the
critical behavior of the dilute Ising model at $p_c$ leading to a
crossover exponent 1 in any $d$. While for a dilute Heisenberg
system the crossover exponent is related to the resistivity
exponent, in agreement with the experimental data of Ref. \cite{BC}.

In conclusion we can write the following relations
\beq
y_H =D_p~~~~~~~y_T =D_R
\label{yh}
\eeq
where $y_H =d - \beta_p/ \nu_p$ is the so-called magnetic field
scaling exponent and $y_T =1/ \nu_p$ is the thermal scaling
exponent in the renormalization group language.
This result is quite interesting as it shows that the scaling exponents can be
expressed in terms of geometrical quantities: The fractal dimension $D_p$ of the
entire incipient infinite cluster, and the fractal dimension $D_R$ of the 
subset made of red bonds.

\subsection{\bf Surfaces and interfaces}
\label{SI}

The study of the structure of the surfaces and interfaces of the
large clusters below $p_c$ has not received as much attention as
the study of the internal structure of the IIC.
This problem is relevant to the study of the dielectric
constant of random composite materials, the viscosity of a gel,
the conductivity of a random superconducting network, and the
relative termite diffusion model.

For simplicity, let us consider a random superconducting network
in which superconducting bonds are present with probability $p$
and normal bonds carrying a unit resistance with probability
$1-p$. For small values of $p$ we have finite superconducting
clusters in a background of normal resistor. As $p \rightarrow
p_c$, the superconductivity $\Sigma$ diverges. For a finite cell of
linear dimension $L$ just
below $p_c$, the typical configurations are characterized by two
very large clusters almost touching, each one attached to one of
two opposite faces. Inside these clusters there are islands of
normal resistors. If a unit voltage is applied between the
opposite face of the hypercube, there is no current flowing
through the bonds in the island. We call these ``dead" bonds, in
analogy with the dead ends of the percolating cluster. The
remaining normal bonds connect one superconducting cluster
to the other. These bonds are made of ``bridges", also called ``antired bonds",
which have the property that if one is
replaced by a superconducting bond, a percolating superconducting
cluster is formed, and
the remaining multiple ``connecting" bonds. Similarly to the
red bonds, it can be proved \cite{coniglio85} that the fractal
dimensionality of the antired bonds is $1/ \nu_p$. The proof is
based on the
following relation which can be proved for any lattice in any dimension
\beq
(1-p) \, \frac{dp_{ij}}{dp}= \mu_{ij}\, ,
\label{lattice}
\eeq
where $p_{ij}$ is the pair connectedness function (the
probability that sites $i$ and $j$ belong to the same cluster)
and $\mu_{ij}$ is the average number of antired bonds between $i$ and
$j$. These are defined as non-active bonds, such that if one is
made active, $i$ and $j$ become connected.

The above considerations suggest that just below $p_c$ the system
can be imagined as a superlattice made of large critical clusters
whose centers are separated by a distance of the order $\xi$.
The surfaces of these clusters almost touch, and are connected by
bridges made of single bonds and other paths made of more than
one bond \cite{CS}.

Finally we mention the following result which relates the size of
the critical cluster $s^*$ and the size of the entire perimeter
$t^*$ \cite{SA,bunde}
\beq
t^* = \frac{1-p}{p} s^* -A s^{* \sigma_p}\, ,
\eeq
where ${\displaystyle \sigma_p = \frac{1}{\nu_p D_p}}$ is the
critical exponent which appears in the cluster number Eq. (\ref{ns}) and
$A$ is a constant. The last term $s^{*\sigma_p}$ which
appears also in Fisher's droplet model \cite{fisher} is usually interpreted as
the surface of the droplet. However,
if it was a surface, $\sigma_p$ should satisfy the following bound
${\displaystyle \frac{d-1}{d} \leq \sigma_p \leq 1}$. The upper
bound corresponds to the fully rarefied droplets and the lower
bound to compact droplets. Surprisingly enough for the
percolation problem $\sigma_p$ is strictly smaller than
${\displaystyle \frac{d-1}{d}}$. This paradox can be solved by
using a result \cite{coniglio83}, which shows that $A s^{* \sigma_p} $ is
equal to number of antired bonds between critical cluster separated by a
distance of order $\xi$. Since the subset of antired bonds is
only a subset of the entire perimeter, it explains why
${\displaystyle \sigma_p < \frac{d -1}{d}}$. This result gives the best
geometrical interpretation of the thermal scaling exponent $y_T$.
It in fact shows that  $y_T = D_{AR}$,
where $ D_{AR}$ is the fractal
dimension of the antired bonds namely
that part of the surface  which contributes to the surface tension.

\section{\bf Percolation in the Ising Model}
\label{CP}
In this section we want to extend the percolation problem to the
case in which the particles are correlated. The simplest model to
consider is the lattice gas or Ising model. In the following we will
use the Ising terminology. We know that Ising model exhibits a
thermodynamic transition for zero external field, $H=0$, at a
critical temperature $T_c$. The question that we ask is how the
percolation properties are modified due to the presence of
correlation. We first consider the case when the cluster are made of
nearest neighbor down spins (Sect 3.1). Later in Sect 3.2 we will
modify the cluster definition in such a way that these new clusters
describe the thermal fluctuations namely we require that the
clusters satisfy the same properties as the droplets in Fisher's
droplet model \cite{fisher}. Namely: i) the size of the clusters
must diverge at the Ising critical points, ii) the linear dimension
of the clusters must diverge with the same exponent as the
correlation length, and iii) the mean cluster size must
diverge with the same exponent as the susceptibility.

These conditions are satisfied if the cluster size distribution for
zero external field has the following form

\beq n(s,T)=s^{-\tau} f((T-T_c)s^\sigma). \eeq

The parameters $\sigma$ and $\tau$ are related to critical exponents
$\alpha$, $\beta$ and $\gamma$ through Eqs. (10) and (11), where now
$\alpha$, $\beta$ and $\gamma$ are the Ising critical exponents.
In particular for $d=2$, $\sigma=8/15\simeq 0.53$ and $\tau=31/15\simeq 2.07$,
and for $d=3$, $\sigma\simeq0.64$ and $\tau\simeq2.21$.

\subsection{\bf Ising Clusters}
\label{CLGIM}
The Hamiltonian of the Ising model is given by:
\begin{equation}
\mathcal{H} = - J \sum_{\langle ij\rangle }S_i S_j -H \sum_i S_i \,
\label{a}
\end{equation}
where $S_i = \pm 1$ are the spin variables, $J$ is the interaction
between two nearest neighbour ($nn$) spins and $H$ is the magnetic
field.

From the thermodynamic point of view the only quantities of
interest are those which can be obtained from the free energy and
those were the only quantities that Onsager was concerned with in
his famous solution of the $2d$ Ising model. However one can look at
the Ising model from a different perspective by studying
the connectivity properties using concepts such as cluster which
have been systematically elaborated in percolation theory
\cite{stauffer1}. There are two reasons for approaching the
problem also from the connectivity point of view. One reason is
that it gives a better understanding of the mechanism of the
phase transition \cite{fisher}. In fact,  concepts like
universality and scaling have been better understood in terms of
geometrical clusters and fractal dimensions \cite{coniglio89}.
A second reason is that there are physical quantities amenable to
experimental observations, which are associated to the
connectivity properties and cannot be obtained from the free
energy. It is very important to note however that the definition
of connectivity, and therefore the definition of the cluster, is
not always the same, but may depend on the particular observable
associated to it.

\begin{figure}[ht]
\hspace{3cm}
\begin{minipage}[t]{0.4\linewidth}
\includegraphics[width=6cm]{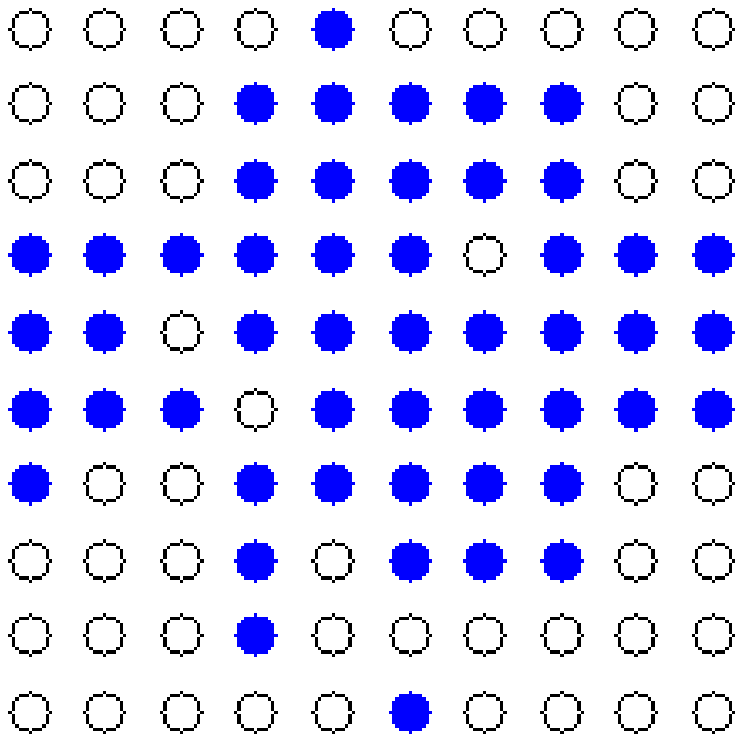}
\end{minipage}%
\hspace{0.1cm}
\begin{minipage}[t]{0.55\linewidth}
\includegraphics[width=6cm]{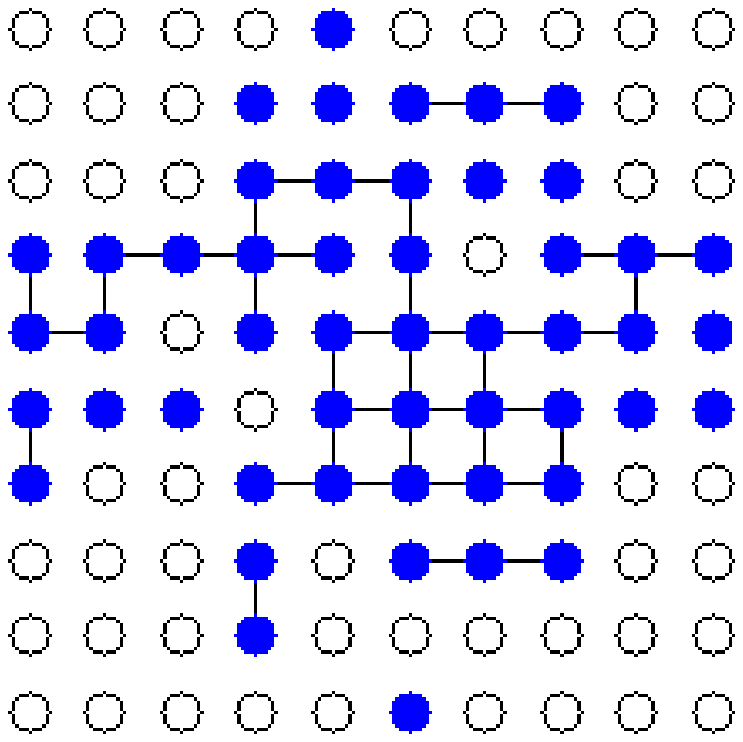}
\end{minipage}%
\caption{(a) Ising configuration at $T_c$:
``down" spins are represented by filled
circles. (b) Correct clusters are obtained from the configuration
given in (a) by putting bonds between occupied sites with probability
$p = 1-e^{-2 \beta J}$.}
\label{coni3}
\end{figure}

In the Ising model, for a given configuration of spins it is rather
natural to define a cluster as a maximal set of $nn$ down parallel
spins \footnote{The Ising Hamiltonian, Eq. (\ref{a}), is equivalent
to the lattice gas Hamiltonian ${\cal H}_{LG} = - J'
\sum_{\langle ij\rangle }n_i n_j -\mu \sum_i n_i$, with
$n_i=(1-S_i)/2$, $J'=4J$ and $\mu=2H-4J$. In the lattice gas
terminology an Ising cluster is a maximal set of $nn$ occupied sites.}
(Fig. \ref{coni3}). For some time these clusters were
believed to be responsible for the correlations present in the Ising
model. This idea was also based on numerical results which showed
evidence that in two dimensions the mean cluster size diverges at
the thermal critical point \cite{binder}. However the idea that the
clusters could describe thermal correlations was definitively
abandoned when it was shown, by numerical simulations in the three
dimensional Ising model \cite{muller} and by exact solution on the
Bethe lattice \cite{conigliobethe}, that the percolation point
appeared in the low density phase of down spins on the coexistence
curve at a temperature $T_{p}$ before the critical point $T_{c}$ is
reached ($T_{p}<T_{c}$).

Later it was suggested by topological arguments
\cite{coniglio3} that only in two dimensions the critical point
coincides with the percolation point, but not necessarily in higher
dimensions. The arguments followed two steps: in the first step it was argued
that an infinite cluster of up spins is a necessary condition for
having a spontaneous magnetization. This implies a percolation
transition of down spins on the coexistence curve $T_p \leq T_c$,
in the second step it was argued that due to topological reasons
in two dimensions it is not possible to have an infinite cluster of
up spins coexisting with an infinite cluster of down spins, which
implies $T_p \geq T_c$. Combining with the previous inequalities
one obtains in two dimensions $T_p = T_c$.

Later these results were proven rigorously \cite{15., 17.} along
with many other results relating connectivity and thermodynamic
quantities. For more details we refer to the original papers.

%

It is clear that the Ising clusters, defined as
group of $nn$ parallel spins, do not have the property of
describing correlated regions corresponding to spin
fluctuations, as originally expected. In fact even in two
dimensions, where the thermal critical point coincides with the
percolation point, the Ising clusters were not suitable for such
description. Series expansion showed that the mean cluster size
diverges with an exponent, $\gamma^* = 1.91 {\pm} 0.001$, rather
different from the susceptibility exponent, $\gamma = 1.75$
\cite{sykes}. Later it has been shown exactly that $\gamma^* =91/48$
\cite{stella}.

\subsection{\bf Ising Droplets}
\label{droplets}

From the properties mentioned in Sect. \ref{CLGIM}, it
appears that the Ising clusters are too big to describe the proper
droplets. The reason is that there are two contributions to the
Ising clusters. One is due to correlations but there is another
contribution purely geometrical due to the fact that 
two $nn$  spins even in absence of correlation
have a finite probability of being parallel. The last contribution
becomes evident in the limit of infinite temperature and zero
external field. In this case in fact, although there is no
correlations and the susceptibility is zero, the cluster size is
different from zero. In fact in $3d$ at infinite temperature there is
even an infinite cluster of ``up" and ``down" spins.

Binder \cite{binder} proposed to cut the infinite cluster in order to
have $T_p = T_c$ in $d=3$, but he did not give the microscopic
prescription to do it. Later Coniglio and Klein \cite{CK} proposed to reduce the
cluster size by introducing fictitious bonds between $nn$ parallel
spins with probability $p_b$ (Fig. \ref{coni3}).
These new clusters are made of $nn$ parallel spins connected by bonds. The
original Ising cluster will either reduce its size or will break into smaller
clusters. If $p_b=1$, we obtain the Ising clusters again. This case is known as
the site correlated percolation problem because one looks at the properties of
the Ising clusters just as in the random percolation problem. The main
difference is that in random percolation the occupied sites are randomly
distributed, while in this case the down (or up) spins are correlated
according to the Ising Hamiltonian. In the infinite temperature limit one
recovers random percolation. The case $p_b\neq 1$ is called site-bond
correlated percolation \cite{csk}.

A Hamiltonian formalism was proposed to study site correlated
percolation \cite{murata}. This formalism was generalized in Ref.
\cite{CK} to study site-bond correlated percolation. In this case
for zero external field the Hamiltonian is given by the following
dilute Ising $s-$state Potts Model (DIPM) \footnote{Originally in
Ref. \cite{CK} the Hamiltonian of the DIPM, $\mathcal{H}_{DP}$,
was expressed in terms of the lattice gas variables $n_i$, and the
Ising droplets were defined as $nn$ occupied sites connected by
bonds, corresponding to $nn$ down spins.}:
\begin{equation}
- \mathcal{H}_{DP}=J_b\sum_{<ij>} (\delta_{\sigma_i\sigma_j}-1)(S_iS_j+1) +
J\sum_{<ij>} S_i S_j,
\end{equation}
where $\sigma_i = 1,\dots, s$ are Potts variables and the sum is
over all nearest neighbor sites. In the same way as the s-state
Potts model in the limit $s=1$ \cite{wu} describes the random bond
percolation model, the DIPM describes percolation in the Ising
model where the clusters are made of parallel spins connected by
bonds with probability, $p_b = 1- e^{-2\beta J_b}$.

In particular the average number of clusters $G$, that plays the
role of the free energy in the percolation problem is given by
$G=dF/ds|_{s=1}$, where
\begin{equation}
-\beta F=\lim_{N\rightarrow \infty}\frac{1}{N} \ln \left(\sum_{\{\sigma_i S_i\}}
e^{-\beta \mathcal{H}_{DP}}\right).
\end{equation}

At that time the DIPM was investigated in a different context by
Berker {\em et} al \cite{berker}. The model exhibits the
interesting properties that by choosing $J_b = J$ it coincides
with a pure $s+1-$state Potts model. Therefore in the limit $s=1$
the DIPM coincides with the $s=2$ Potts model namely with the
Ising model. Consequently $F$ becomes the Ising model free energy
and $G$ has a singularity at the Ising critical point. This
argument immediately suggested that the site-bond correlated
percolation for $J_b = J$ namely with the bond probability given
by
\begin{equation}  \label{pbond}
p_b \equiv p = 1 - e^{-2\beta J},
\end{equation}
should reproduce the same critical behavior of the Ising model.
Namely the percolation quantities become critical at the Ising
critical point in the same way as the corresponding thermal
quantities.

In fact using real space renormalization group arguments, it was
possible to show that
the size of the clusters of parallel spins connected by bonds  with
probability, $p_b$, given by Eq. (\ref{pbond}),
diverges at the Ising critical point with Ising exponents,
exhibiting thus the same properties as the droplets in Fisher's
model. These clusters were called droplets to
distinguish them from the Ising clusters.

\subsection{\bf Droplets in $2$ and $3$ dimensions}
\label{2d-3d}

This site-bond correlated percolation problem has been
studied by real space renormalization group
in two dimensions \cite{CK,zia}, by $\epsilon$ expansion, near six
dimensions \cite{ConLub} and by Monte Carlo in two and three
dimensions \cite{stauffer1,Roussenq,Jan,odagaki,Heermann}.

The renormalization group analysis shows that in $2d$ the Ising
critical point is a percolation point for down or up spins connected
by bonds for all values of bond probability such that $ 1 \le p_b <
1-e^{-2 \beta J}$. The fractal dimension $D^*= (\gamma^*/\nu + 2)/2=187/96$
\cite{stella} 
being  higher than the fractal dimension $D = (\gamma/\nu + 2)/2=15/8$
for the value of $p_b \equiv p=1 - e^{-2\beta J}$.

In the renormalization group language this means that there are two
fixed points, one corresponding to the universality class of the
Ising cluster, the other one corresponding to the droplets. In the
first one the variable $J_b$ is irrelevant namely the scaling
exponent associated to it, $y_b<0$. In the second fixed
point associated to the droplets instead $y_b>0$. The result that
the Ising critical point is a percolation point for a range of
values of $p_b$, at the first sight seems counter-intuitive. In fact
if the Ising critical point corresponds to the onset of percolation
for Ising clusters ($p_b = 1$),  one would expect that for $p_b < 1$
the clusters would not percolate anymore. The puzzle can be
clarified by studying the fractal structure of the Ising clusters
and the droplets at $T_c$ \cite{coniglio89}. In fact it can be shown
that $y_b$ is the scaling exponent of the red bonds, namely
$L_R\sim l^{y_b}$ where $L_R$ is the number of red bonds between two
connected sites separated by a distance of the order $l$,
consequently the droplets, characterized by $y_b>0$, are made of
links and blobs, like in random percolation. Due to the presence of
links the cluster breaks apart and does not percolate anymore as the
bond probability decreases. On the contrary the Ising clusters ($p_b = 1$), 
characterized by  $y_b<0$,  are made only of blobs 
and no links, therefore
by decreasing the bond probability the infinite cluster does not break and still
percolates, until $p_b = p$.

In $3d$ at the Ising critical point, $T_c$, there is an analogous line
of anomalous percolation point for clusters of down spins connected
by bonds, for all values of bond probability such that $1\le p_b < 1 -
e^{-2\beta J}$, although the probability $P_{\infty}$ for a down spin
to be in the infinite cluster is different from zero. More precisely
the quantity $p_{ij} -P^2_{\infty} $ decays as a power law, where
$p_{ij}$ is the probability that $i$ and $j$ are connected. For more
details see Ref. \cite{figari}. As $p_b$ decreases towards $p= 1 -
e^{-2\beta J}$ there is a crossover towards a different power law
characterized by the Ising exponent, while $P_{\infty} $ goes to
$0$.

\subsection{\bf Droplets in an external field}
\label{ex_field}

\begin{figure}[ht]
\begin{center}
\epsfig{file=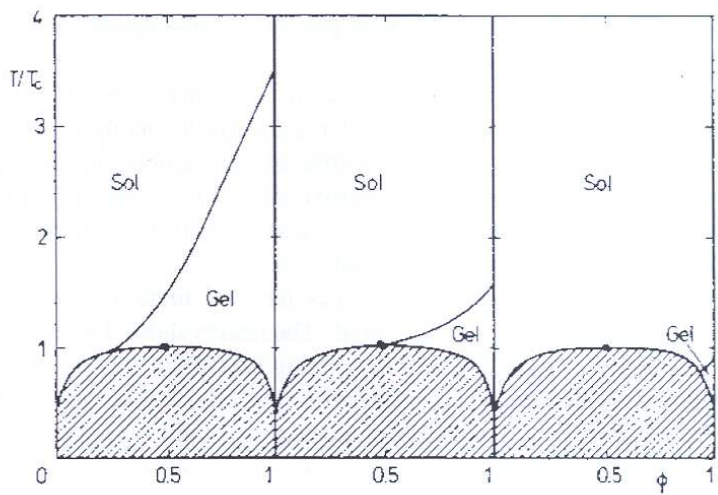, height=7cm}
\end{center}
\caption{\label{5} Montecarlo simulations of
the $3d$ lattice gas model for three values of the bond probability
$p_b = 1-e^{-2 c \beta J}$ with the constant $c =
2.25,\, 1,\, 0.564$ from left to right. $\Phi$ is the density of down spins.
The Gel and the Sol
indicates the percolation and non 
percolation phase. From
Ref. \cite{KCS}.}
\end{figure}

By keeping the same definition of droplets given above, in the case of an
Ising model in an external field $H>0$ one finds a phase diagram in the $H,T$
plane or in the $M,T$ plane, with a percolation line of ``down" spins ending
at the Ising critical point (see Fig. \ref{5}).
Along the percolation line one finds critical exponents in the
universality class of random percolation with a cross-over to Ising
critical exponents as the Ising critical point is approached \cite{CK}. In this context the
Ising critical point is a higher order critical point for the
percolation transition. This percolation line, also known as the
Kertesz line, has received some attention
\cite{SA,Kertesz89,stauffer90,wang89} (see for more details the
review by Sator \cite{sator}).
Although the Ising free energy has no
singularity along this line some physical interpretation is given to
the Kertesz line \cite{campisator}.

On the other hand this line disappears if the droplet definition is
modified in the presence of an external field
\cite{conigliodiliberto, wang89}, according to Kasteleyn and
Fortuin formalism \cite{3} and Swendsen and Wang approach \cite{SW87}.
In this approach
the field is treated as a new interaction between each spin and a
ghost site.
Consequently for positive $H$ (negative $H$)  an ``up" (``down") spin can
be connected to the ghost spin with a probability
$p_H = 1 - e^{-2\beta |H| }$. Droplets now are defined as a maximal
set of spins connected by bonds where as before the bonds between
nearest neighbor parallel spins have probability $p_b$ given by Eq.
(\ref{pbond}) and $p_H$ between spins and the ghost spin. Note that
two far away spins can be easily connected through the ghost spin.
In this way the presence of a positive (negative) magnetic field
implies always the presence of an infinite cluster of ``up"
(``down") spins.

\subsection{\bf Exact relations between connectivity and thermal properties}
\label{exact}
Interestingly it was also shown \cite{conigliodiliberto} that the
droplets so defined with and without the external field
have the same statistics of the clusters in
the random cluster model introduced by Kasteleyn and Fortuin
(KF) \cite{3} (see Sect. \ref{FKRCM}),
although the CK droplets and the KF clusters have a different
meaning. Using the relations between the connectivity properties of
the random cluster model and the thermal properties of the Ising
model, it was finally possible to prove that in any dimension and
for any temperature $T$ and external field $H\ge0$,
the following relations between connectivity and 
thermal properties hold \cite{conigliodiliberto}:
\begin{equation}
\left\{\begin{array}{l}
\rho_{\infty} = m \\
~~ \\
p_{ij} = g_{ij}
\end{array}\right.
\label{pij}
\end{equation}
where $\rho_{\infty}$ is the density of up spins in the
percolating droplet, $m$ is the magnetization per site, $p_{ij}$
is the probability that $i$ and $j$ are connected (through both
finite or infinite droplet) and $g_{ij} =<S_iS_j>$.

In particular, for $T>T_c$ and zero external field $H\rightarrow 0$,
we have that the magnetization  $m=0$ and $g_{ij}$ coincides with
the spin-spin pair correlation function. Consequently
$\rho_{\infty}=0$, namely the probability for a spin to be in an
infinite droplet is zero, and therefore $p_{ij}$ coincides with the
probability that two spins $i$ and $j$ are in the same finite droplet.
For $T<T_c$ instead we have $\rho_{\infty} = m>0 $,
and $p_{ij}=p^f_{ij} + p^{\infty}_{ij}$, where $p^f_{ij}$
($p^{\infty}_{ij}$) is the probability that spins in $i$ and $j$ are
in a finite (infinite) droplet. From Eq. (\ref{pij}) it follows for
$T<T_c$:
\begin{equation}
 p^f_{ij} + p^{\infty}_{ij} -
\rho^2_{\infty} = <S_iS_j> - m^2.
\label{pij2}
\end{equation}

By summing over $i$ and $j$ we have
\begin{equation}
S + (\Delta \rho_{\infty})^2 =\chi,
\label{chi}
\end{equation}
where $S$ is the mean cluster size of the finite clusters, 
$(\Delta \rho_{\infty})^2$ is the
fluctuation of the density of the infinite cluster and $\chi$ is the
susceptibility. These exact results show that above $T_c$ mean
cluster size and susceptibility coincide, while below $T_c$ there
are two contributions to the susceptibility, one due to the mean
cluster size and the second related to the fluctuation of the
density of the infinity cluster. Monte Carlo calculations \cite{Roussenq}
show that both term have the same critical behavior as also occurs
in random percolation \cite{con_stau,chayes1}, so the mean cluster
size $S$ diverges like the susceptibility. We expect that this is the
case for dimensions up to $d=4$, the upper critical dimensionality of
the Ising model. In mean field, as we will see, the mean cluster size
below $T_c$ diverges with an exponent different from the
susceptibility.


One very interesting application based on the KF approach was
produced by Swendsen and Wang \cite{SW87,WS90}, who elaborated a
cluster dynamics which drastically reduced the slowing down near the
critical point of the Ising and Potts model (see also \cite{Wolff}
for further developments).

The droplet definition can be extended to the $nn$ antiferromagnetic
Ising model \cite{amitrano} and to the Ising model with any ferromagnetic 
interaction $J_{ij}$ between sites $i$ and $j$ \cite{Jan}. 
In this case the CK clusters are defined as set of
parallel spins connected by bonds present between $i$ and $j$ with probability
$p_{ij}=1-\exp{[-2\beta J_{ij}]} $.
It can be
shown that also in this case the relations (\ref{pij})
between connectivity and thermal quantities hold.

%

\subsection{\bf Ising Droplets above $d=4$}
\label{IDAD}
In Sect. \ref{droplets} we have reported the relations Eq.s (\ref{pij})
and (\ref{pij2}),
which are exact and are valid in any dimension including mean field.
As a matter of fact in mean field the percolation order parameter
and the magnetization are identical and go to zero with the exponent
$\beta = 1/2$, while the mean cluster size above $T_c$ coincides
with the susceptibility and diverges with the exponent $\gamma =1$.
The same is true for the connectedness length above $T_c$, which
coincides with  the correlation length, and diverges with an exponent
$\nu= 1/2$. However below $T_c$ the mean cluster size diverges with an
exponent $\gamma' = 1/2$ and the correlation length with an
exponent $\nu' = 1/4$ \cite{csk,conigliodiliberto,chayes}. This result is a consequence that the two
terms in Eq. (\ref{pij2}), the probability that two sites are in the same
finite droplet,  $p^f_{ij}$, and the correlation of the infinite droplet density
at site $i$ and $j$,  $p^{\infty}_{ij} - \rho^2_{\infty}$, do not scale in the 
same way, giving rise to two
lengths, diverging respectively with exponents $\nu'$ and $\nu$.

These somehow anomalous results are probably a consequence that the
Ising model has an upper critical dimension $d_c=4$ while the DIPM
which describes the droplet problem has an upper critical dimension
$d_c = 6$ \cite{ConLub}. In fact there are arguments that for $4
\leq d \leq 6$ below $T_c$ the critical exponents are
${\displaystyle \nu' =\frac{1}{d-2}}$, ${\displaystyle \gamma'
=\frac{2}{d-2}}$, $\beta =1/2$, $\eta =0$ and fractal dimension $D_p
={\displaystyle \frac{1}{2} (d+2)}$, with an upper critical
dimension $d_c =6$. Of course for $T>T_c$ the exponents are $\gamma
=1$, $\nu= 1/2$ and $\eta =0$.

Due to the breakdown of the mapping between thermal fluctuations and
mean cluster size below $T_c$ above $d=4$, it is not possible to extend easily
the geometrical picture, employed  in random percolation, 
to explain the breakdown of hyperscaling  in the Ising
model. For a study of droplets inside the metastable region
see Ref. \cite{klein_k}.

\subsection{\bf Generalization to the $q$-state Potts Model}
\label{potts}
All the results found for the Ising case have been
extended \cite{CP} to the $q$-state Potts model. This model is
defined by the following Hamiltonian:
\begin{equation}
- {\cal H}_q=qJ\sum_{<ij>}\delta_{\sigma_{i}\sigma_{j}},
\end{equation}
where the spin variables $\sigma_{i}$ can assume $q$ values,
$\sigma_{i}=1,\dots,q$. This model coincides with the Ising model
for $q=2$, reproduces the random percolation problem in the limit
$q=1$ and the tree percolation model in the limit $q=0$ \cite{wu}.
The geometrical approach developed in the previous sections for
the Ising model, can be extended to the $q$-state Potts model. In
particular one can define the site-bond Potts correlated
percolation, where clusters are made of $nn$ spins in the same
state,
connected by bonds with bond probability $p_b$. By choosing $p_b
=p=1-e^{-q\beta J}$, it is possible to show that these clusters
percolate at the Potts critical temperature $T_c(q)$, with
percolation exponents identical to the thermal exponents and
therefore behave as the critical droplets.


The formalism is based on the following diluted Potts model
\cite{CP,varie}:
\begin{equation}
- {\cal H}^q_{DP}=J_b\sum_{<ij>} (\delta_{\tau_i\tau_j}-1)
\delta_{\sigma_{i}\sigma_{j}} + q J\sum_{<ij>}
\delta_{\sigma_{i}\sigma_{j}},
\end{equation}
where the second term, which controls the distribution of spin variables, is
the $q-$state Potts Hamiltonian, whereas the first term contains auxiliary
Potts variables $\tau_i=1,2,\dots,s$ and controls the bonds distribution.

As in the Ising case, Hamiltonian (37) in the limit $s \rightarrow 1$
describes the site-bond Potts correlated percolation problem with
$p_b$ given by $p_b=1-e^{-q\beta J_b}$. The droplets are obtained
in the particular case $J_b=qJ$. For this value in fact
Hamiltonian (37) for $s \rightarrow 1$ coincides with the $q$-state
Potts model.



Once the Ising and Potts model has been mapped onto a percolation
problem, we can extend some of the results of random percolation
to thermal problems.

\subsection{\bf Fractal structure in the Potts Model: Links and blobs}
\label{links_blobs} 
Like in random percolation, also in the
$q$-state Potts model it can be shown that at $T_c(q)$ the
critical droplets have a fractal structure made of links and
blobs, with a fractal dimension $D(q) = d - \beta (q) / \nu (q)$,
where $\beta (q)$ and $\nu (q)$ are respectively the order
parameter and correlation length exponent. Therefore  $D(q)$
coincides with the magnetic scaling exponent $y_H(q)$. However the
fractal dimension of the red bonds $D_R(q)$ does not coincide with
the thermal scaling exponent $y_T(q)$, associated to the thermal
variable $J$, like in random percolation. Instead $D_R(q)$ is
found to coincide with the bond probability scaling exponent $y_b$
associated to the variable $J_b$ in Hamiltonian (37)
\cite{coniglio89}.

Like for random percolation the fractal dimension of the red bonds
coincides with the fractal dimension of the antired bonds. Using
the mapping from the Potts model to the Coulomb gas
\cite{duplantier}, it is possible to obtain the exact value of the fractal
dimension
of the red bonds and of the external perimeter or hull
\cite{coniglio89}. For further exact results see also
\cite{B1pg16,stella}.

\begin{table}
\caption{\label{tab1} Fractal dimensions, for $d=2$, of the whole cluster ($D$),
of the Hull ($D_H$), and of the red bonds ($D_R$) for the Potts
droplets. It is also reported the thermal power exponent $y_T$.}
\begin{tabular*}{\textwidth}{@{}l*{15}{@{\extracolsep{0pt plus
12pt}}l}} \br
$q$&$D$&$y_T$&$D_H$&$D_R$\\
\mr
$0$&$2$&$0$&$2$&$5/4$\\
$1$&$91/48$&$3/4$&$7/4$&$3/4$\\
$2$&$15/8$&$1$&$5/3$&$13/24$\\
$3$&$28/15$&$6/5$&$8/5$&$7/20$\\
$4$&$15/8$&$3/2$&$3/2$&$0$\\
\br
\end{tabular*}
\end{table}
From Table \ref{tab1} it appears that the exact value of $D(q)$
does not vary substantially with $q$, for $d=2$. This observation
can be understood by noting that, using this geometrical approach,
the driving mechanism of the critical behavior can be viewed as
coalescence of clusters just like in random percolation.
Then one would expect for any $q$ that the fractal dimension
should be close to the fractal dimension of the critical clusters
in the percolation problem. This also explains the observation of
Suzuky \cite{suzuki}, known as strong universality, that for a
large class of models the ratio $\gamma /\nu $ or $\beta /\nu $ do
not vary appreciably. Since these ratios of critical exponents for
fixed $d$ depend only on the magnetic scaling exponent, which is
identical to the fractal dimension, the strong universality is
consequence of the quasi-universal feature of the fractal
dimension as discussed above. Unlikely the fractal dimension of
the whole cluster, $D_R(q)$ and $D_H(q)$ do change substantially
and characterize the different models as function of $q$.
Particularly sensitive to $q$ is the fractal dimension of the red
bonds, which has its largest value at $q = 0$ (tree percolation),
where the backbone is made only of links. As $q$ approaches $q_c$
the cluster becomes less ramified until the red bonds vanish
($D_R(4) = 0$). This results in a drastic structural change from a
links and blobs picture to a blobs picture only, anticipating a
first order transition. Interestingly, the fractal dimension of
the red bonds for $q=0$, $D_R=5/4$, has been related to the
abelian sandpile model \cite{dhar}. The reason why $D_R(q)$ is so
model dependent is due to the fact that the fractal set of the red
bonds is only a small subset of the entire droplet, and therefore
this ``detail" is strongly model dependent. Also the thermal
exponent $y_T(q)$ is strongly model dependent, however so far it
has not been found the geometrical characterization in terms of a
fractal dimension for such exponent except $q=1$ (random
percolation).

\subsection{\bf Fortuin Kasteleyn-Random Cluster Model}
\label{FKRCM} 
We will present  here the random cluster model
introduced by Kasteleyn and Fortuin. Let us consider the q-state
Potts model on a $d-$dimensional hypercubic lattice. By freezing
and deleting each interaction of the Hamiltonian (see Appendix),
they managed to write the partition function of the Potts model,
$Z=\sum_{\{\sigma_i\}} e^{-\beta {\cal H}_q}$, in the following
way
\begin{equation}
Z = \sum_C p^{|C|}(1-p)^{|A|} q^{N_C} \, ,  
\label{zkf_bis}
\end{equation}
where $C$ is a configuration of bonds defined in the same hypercubic
lattice, just like a bond configuration in the standard percolation
model, $|C|$ and $|A|$ are respectively the number of bonds present
and absent in the configuration $C$, and $N_C$ is the number of
clusters in the configuration $C$.

In conclusion, in the KF formalism the partition function of the
Potts model is identical to the partition function (\ref{zkf_bis}) of a
correlated bond percolation model \cite{3, hu} where the weight of
each bond configuration $C$ is given by
\begin{equation}
W(C) = p^{|C|}(1-p)^{|A|} q^{N_C}
\label{wkf1_bis}
\end{equation}
which coincides with the weight of the random percolation except for
the extra factor $q^{N_C}$. They called this particular correlated
bond percolation model, the random cluster model. Clearly for q = 1
the the cluster model coincides with the random percolation model.


Kasteleyn and Fortuin have related the percolation
quantities associated to the random cluster model to the
corresponding thermal quantities in the $q-$state Potts model \cite{3}.
In particular for the Ising case, $q=2$,
\begin{equation}
| \langle S_i \rangle | = \langle \gamma_i^{\infty}\rangle_{W}
\label{new1}
\end{equation}
and
\begin{equation}
\langle S_iS_j\rangle=\langle\gamma_{ij}\rangle_{W} \, ,
\label{new2}
\end{equation}
where $\langle ... \rangle$ is the Boltzmann average and 
$\langle ...\rangle _{W}$ is the average over bond configurations in the bond
correlated percolation with weights given by (\ref{wkf1_bis}). Here $
\gamma_i^{\infty}(C)$ is equal to 1 if the spin at $i$ belongs to
the infinite cluster, $0$ otherwise; $\gamma_{ij}(C)$ is equal to 1
if the spins at sites $i$ and $j$ belong to the same cluster, $0$
otherwise. 

Interestingly the connectivity properties in the KF random cluster model can
be related to the CK droplets:  
\begin{equation}
\rho_\infty=\langle \gamma_i^{\infty}\rangle_{W},
\label{new3}
\end{equation}
\begin{equation}
p_{ij}=\langle\gamma_{ij}\rangle_{W},
\label{new4}
\end{equation}
where $\rho_\infty$ and $p_{ij}$ are defined in Sect. \ref{exact}.
From Eq.s (\ref{new1}-\ref{new4}) it follows Eq.s (\ref{pij}). 
 
%

\section{\bf Hill's clusters}
\label{secthill}

In this section we discuss the possibility to extend the definition of
droplets to simple fluids. In 1955 Hill \cite{hill} introduced the concept of
physical clusters in a fluid in an attempt to explain the
phenomenon of condensation from a gas to a liquid. In a fluid
made of particles interacting via a pair potential $u(r)$
physical clusters are defined as a group of particles pairwise
bounded. A pair of particles is bounded if in the reference frame
of their center of mass their total energy is less than zero.
Namely their relative kinetic energy plus the potential energy is
less than zero. The probability that two particles at distance
$r$ are bounded can be calculated \cite{hill} and is given by
\begin{equation}
p_{\, \mbox{\tiny H}} (r) = \frac{4}{\pi} \int_0^{\sqrt{-\beta u
(r)}} x^2 e^{-x^2}\, dx \, . \label{phill}
\end{equation}

More recently it was noted \cite{campi} that the bond probability
Eq. (\ref{phill}) calculated for the interaction of the three
dimensional $nn$ lattice gas model is almost coincident with the
bond probability $p$ of Eq. (\ref{pbond}).
This implies that
Hill's physical clusters for the $3d$ lattice gas almost coincide
with the droplets defined by Coniglio and Klein, and in fact
Hill's clusters percolate along a line almost indistinguishable
from the droplets percolation line (see Fig. \ref{5}).

In order to calculate percolation quantities in a fluid in Ref.
\cite{deangelis} the authors developed a theory based on Mayer's
expansion. In particular, using this theory they calculated
analytically for a potential made of hard core plus an attractive
interaction, the percolation line of Hill's physical clusters in a
crude mean field approximation and compared with the liquid gas
coexistence curve. They found that the percolation line ended just
below the critical point in the low density phase but not exactly
at the critical point. For further developments of the theory see
\cite{Gpag17}.

\begin{figure}[ht]
\begin{center}
\epsfig{file=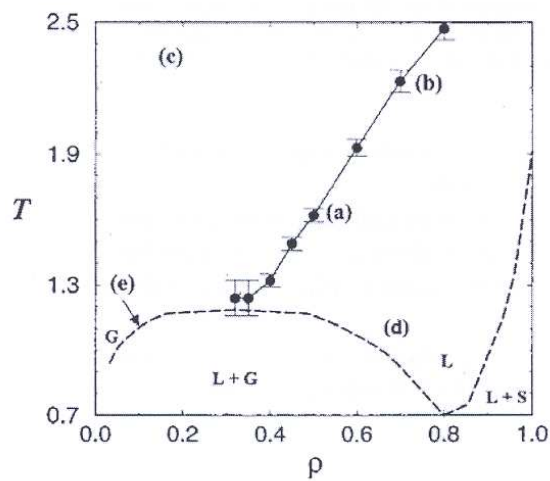, height=7cm}
\end{center}
\caption{\label{fig6} Phase diagram of the Lennard--Jones fluid
using molecular dynamics. The full line corresponds to
percolation of cluster following Hill's definition. From Ref.
\cite{campisator}.}
\end{figure}

Very recently, Campi {et al.} \cite{campisator}, using molecular
dynamics have calculated the percolation line of Hill's physical
clusters for a Lennard--Jones potential. The results showed a
percolation line ending close or at the critical point (Fig. \ref{fig6})
suggesting that Hill's clusters are good candidates to describe
the density fluctuations like the droplets in the lattice gas
model, although there is no proof of relations analogous to those
valid for the droplets in the lattice gas such as Eq.
(\ref{pij}), which would prove that their size would diverge
exactly at the critical point with thermal exponents.

Although Hill's clusters may represent the critical fluctuation
near the critical point, we may wonder whether they have a
physical meaning away from the critical point. In particular we
may wonder whether we can detect experimentally the percolation
line in the phase diagram. In a Lennard--Jones fluid, molecular
dynamics shows that quantities such as viscosity or diffusion
coefficient do not seem to exhibit any anomalous behaviour
through the percolation line \cite{campisator}. In some colloids
instead the percolation line is detected through a steep increase
of the viscosity. What would be the difference in the two cases?
The difference may rely in their lifetime. The possibility to
detect the percolation line of these clusters is expected to
depend on the lifetime of the clusters which in turn depends on
the bond lifetime. The larger is the cluster lifetime the larger
is the increase of the viscosity, the better the percolation line
can be detected. In Sect. \ref{sect_viscosity} we will discuss the behaviour of
the viscosity as function of the lifetime of the clusters.

\section{\bf Clusters in weak and strong gels}
\label{gel}

In the previous section we have shown the case in which the
probability of having a bond between two particles coincides with
the probability that the two particles form a bound state defined
according to Hill's criterion. Now we want to show another
mechanism leading to the formation of bound states, which is more
appropriate to gels. The importance of connectivity in gels was
first emphasized by Flory \cite{flory}. The application of
percolation theory to gels was later suggested by de Gennes
\cite{degennes} and Stauffer \cite{stauffer3, 41.}. Here we
consider a system made of monomers in a solvent. Following
Ref. \cite{csk} we shall assume that the monomers can interact
with each other in two ways. One is the usual van der Waals
interaction, and the other is a directional interaction that
leads to a chemical bond. A simple model for such a system is a
lattice gas model where an occupied site represents a monomer and
an empty site a solvent. For simplicity we can put equal to zero
the monomer-solvent interaction and the solvent-solvent
interaction, and include such interaction in an effective monomer-monomer
interaction. The monomer-monomer interaction
$\varepsilon_{ij}$ can reasonably be approximated by a nearest
neighbour interaction
\begin{equation}
\varepsilon_{ij} = \left\{
\begin{array}{c}
-W \\~~ \\ -E \end{array} \right. \label{epsilonij}
\end{equation}
where $-W$ is the van der Waals type of attraction and $-E$ is the
bonding energy. Of course, this second interaction, which is the
chemical interaction, occurs only when the monomers are in
particular configurations. For simplicity we can suppose that
there is $1$ configuration which corresponds to the interaction of
strength $E$, and $\Omega$ configurations which corresponds to the
interaction of strength $W$. We expect $E \gg W$ and $\Omega \gg
1$. It can be easily calculated \cite{csk} that such a system is
equivalent to a lattice gas model with an effective $nn$
interaction $-\varepsilon$ given by
\begin{equation}
e^{\beta \varepsilon} = e^{\beta E} + \Omega e^{\beta W} \, .
\label{epsilon}
\end{equation}
Therefore from the static point of view the system exhibits a
coexistence curve and a critical temperature which characterizes
the thermodynamics of the system. However the system
microscopically behaves rather different from a standard lattice
gas. In fact in a configuration in which two monomers are $nn$, in
a standard lattice gas they feel one interaction, while in the
system considered here with some probability $p_b$ they feel a
strong chemical interaction $-E$ and with probability $1-p_b$
they feel a much smaller interaction $-W$. The probability $p_b$
can be easily calculated and is given by
\begin{equation}
p_b = \frac{e^{\beta E}}{e^{\beta E} + \Omega e^{\beta W}}\, .
\label{pb}
\end{equation}

In conclusion, the system from the static point of view is
equivalent to a lattice gas with interaction $\varepsilon$ given
by (\ref{epsilon}). However we can also study the percolation line
of the clusters made by monomers connected by chemical bonds. This
can be done by introducing bonds between $nn$ particles in the
lattice gas with $nn$ interactions, the bonds being present  with
probability $p_b$ given by Eq. (\ref{pb}). By changing the solvent
the effective interaction $W$ changes and one can realizes three
cases topologically similar to those of Fig. \ref{5}, where the
percolation line ends at the critical point or below the critical
point in the low density or high density phase (for more details
see \cite{csk}).

The lifetimes of the bonds are of the order of $e^{\beta E}$. Since
$E$ is very large the lifetime could be very large. For an infinite
bond lifetime the bonded clusters are permanent and the viscosity
diverges due to the divergence of the mean cluster size  (see for example
\cite{41.}), and the percolation line can be easily detected. We
consider three particular physical systems which could be rather
emblematic of a general situation where the percolation line has
been detected:
\begin{itemize}
\item[a)]  Microemulsions of water in oil \cite{chen}.

\item[b)]  Triblock copolymers in unicellar systems
\cite{mallamace1, mallamace2}.

\item[c)]  Gelatin water methanol systems
\cite{tanaka}.
\end{itemize}

\begin{figure}[ht]
\begin{center}
\epsfig{file=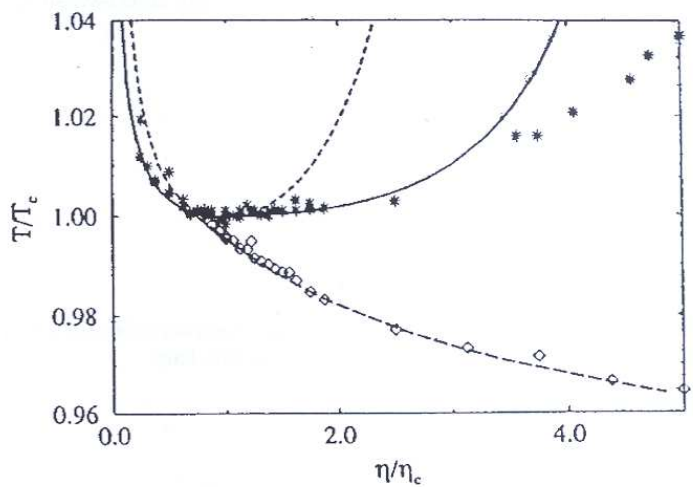, height=7cm}
\end{center}
\caption{\label{1} Experimental points in AOT/water/decane from
Ref. \cite{chen} together with the coexistence curve and spinodal
curve based on the Baxter's model. The percolation line where the
conductivity  exhibits a steep increase has been fitted with the
Baxter's model. }
\end{figure}

\begin{figure}[ht]
\begin{center}
\epsfig{file=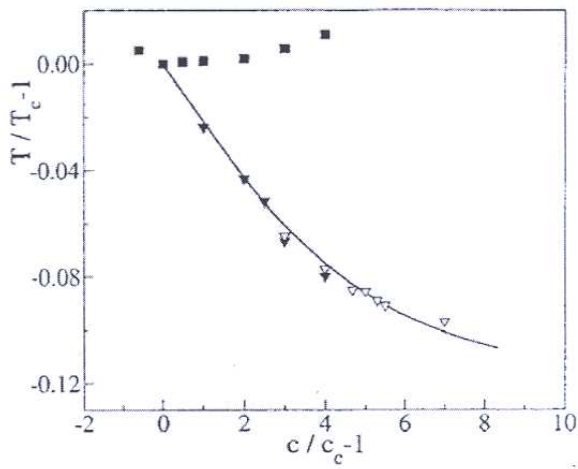, height=7cm}
\end{center}
\caption{\label{2} $L/64$ water system. Experimental points of
the coexistence curve and percolation line, where the viscosity
exhibits a steep increase. From Ref. \cite{mallamace1}. }
\end{figure}

\begin{figure}[ht]
\begin{center}
\epsfig{file=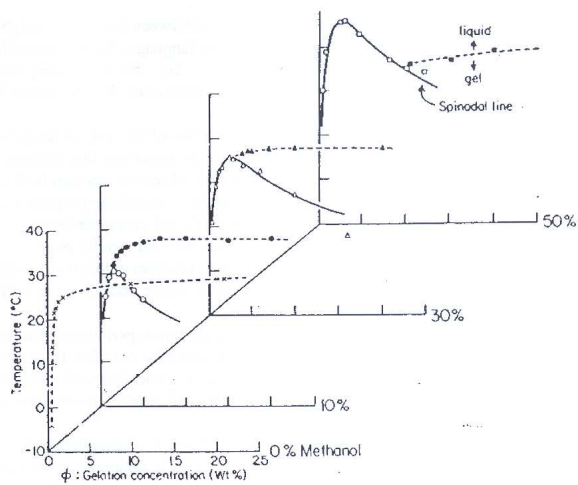, height=7cm}
\end{center}
\caption{\label{3} Sol-Gel transition temperature (solid symbols)
and the spinodal temperature (open symbols) of
gelatin-water-methanol mixtures as function of gelatin
concentration. At the sol-gel transition the viscosity diverges.
From Ref. \cite{tanaka}.}
\end{figure}

In Fig.s \ref{1}, \ref{2} and \ref{3} we show respectively the phase
diagram of the systems a), b), c), where it is shown the coexistence
curve in the temperature-concentration diagram, together with
``percolation lines".

In particular, in a) the system consists of three components
AOT/water/decane. For the temperature and the concentration of
interest, the system can be considered as made of small droplets of
oil surrounded by water in a solvent. The droplets interact via a
hard core potential plus short range attractive interaction. Because
of the entropic nature of the attractive interaction, the
coexistence curve is ``upside-down'' with the critical point being
the minimum instead of the maximum (Fig. \ref{1}). The broken line
is characterized by a steep increase of conductivity.

In b) the system is made of triblock copolymers unicellar in water
solution, $c$ is the volume fraction of the unicelles (Fig.
\ref{2}). The line is characterized by a steep increase in the
viscosity.

In c) the system is made of gelatin dissolved in water + methanol;
$\phi $ is the gelatin concentration. The broken lines are
characterized by the divergence of the viscosity and correspond to
the sol-gel transition. Each line represents a different value of
the methanol concentration, which has been chosen in such a way that
the line ends  at the consolute point or, below it, in the low or
high density phase (Fig. \ref{3}).

In all these experiments the consolute point is characterized by a
thermodynamical singularity, where the correlation length  and
compressibility diverge. The other lines are usually ascribed to a
``percolation'' transition. However it is important to precise
which are the relevant clusters in the three different systems.
Also we would like to understand why, in system c),  the viscosity
diverges at the percolation transition, while in b) it reaches a
plateau, and why in a) and b)  the ``percolation" lines  end on
the coexistence curve close to the critical point in the low
density region . It is also important to realize that for each
phenomenon is very important to define the proper cluster, which
is responsible for the physical phenomenon. In the conductivity
experiments in microemulsion the proper clusters are made of
``touching" spheres similar to nearest neighbor particles in a
lattice gas model.
The viscoelastic properties of microemulsions may be more suitably
described by clusters made of spheres pairwise bonded. 
For a more refined percolation model in microemulsion, see \cite{Grest1986}.

From the cluster properties of the lattice gas model we expect the
infinite cluster is a necessary condition for a critical point
therefore the percolation line ends just below the critical point
in the low density region, as observed in the experiments described above
and more recently in numerical simulations of models of
interacting colloids \cite{int_col2}.


In weak reversible gelatin the clusters are made of monomers (or
polymers) bonded by  strong interaction which leads to chemical
bond. In this case the bond probability can be changed by changing
the solvent and therefore the percolation line, by properly
choosing the solvent, can end on the coexistence curve at or below
the critical point.

The reason why the viscosity in the gel experiments diverges at
the percolation point, while it reaches a plateau in colloids, is
due to the lifetime of the bonds which is much longer in the first
system than in the second  \cite {delgado,sciortino,
mallamace}. In low density colloids the proper cluster to
describe colloidal gelation also appear to be related to strong
bonds with large bond lifetime \cite{noi,bartlett,sciortino}. When
the relaxation time is much smaller than the bond lifetime the
dynamics is dominated by the clusters, otherwise a crossover is
expected towards a regime due to the crowding of the
particles \cite{noi}. Percolation line of clusters pairwise bonded
can also be defined in fluids, but due to the negligible lifetime
cannot be detected.


\section{\bf Scaling behaviour of the viscosity}
\label{sect_viscosity}

If the lifetime of the chemical bonds is infinite, the viscosity
exhibits a divergence at the percolation threshold as recently
shown in different models \cite{180, zippelius, pliske}
\begin{equation}
\eta \sim \xi^{\tilde{k}}  \, ,\label{viscosity}
\end{equation}
where $\xi$ is the linear dimension of the critical cluster which
diverges at the percolation threshold with the exponent $\nu$.

The relation between the diffusion coefficient $D(R)$ of a
cluster of radius $R$ and the viscosity $\eta$ would be given by
the Stokes--Einstein relation for a cluster radius much larger
than $\xi$
\begin{equation}
D (R) \sim \frac{1}{R\eta}\, .\label{stokes}
\end{equation}
For cluster radius $R$ \ smaller than $\xi$ it has been proposed
\cite{martin} that the viscosity will depend also on $R$ in such
a way to satisfy a generalized Stokes--Einstein relation Eq.
(\ref{stokes}) with $\eta = \eta (R)$. When $R = \xi$ the
viscosity $\eta (\xi) = \eta$, and from Eq. (\ref{stokes}) one obtains
the following scaling behaviour for $R$:
\begin{equation}
D(R) \sim R^{-(1 + \tilde{k})} \label{DR}
\end{equation}
therefore the relaxation time $\tau(R)$ for a cluster of radius
$R$ is
\begin{equation}
\tau (R) \sim R^{1 + \tilde{k}} \, .\label{tauR}
\end{equation}
If $\tau$ is the lifetime of a typical cluster, then a cluster of
radius $R$ will contribute to the viscosity if $\tau (R) < \tau$,
and therefore:
\begin{equation}
\eta \sim \xi^{\tilde{k}}f  \left( \frac{\tau}{\xi^{1 +
\tilde{k}}} \right) \sim \left\{ \begin{array}{ll} \xi^{\tilde{k}}
~~~~~ & \tau >
\xi^{1+ \tilde{k}} \\
& \\
\tau^{\frac{\tilde{k}}{1 + \tilde{k}}} & \tau < \xi^{1 +
\tilde{k}}
\end{array} \right. \label{tau}
\end{equation}
which implies that the viscosity will exhibit a steep increase
followed by a plateau. The higher is $\tau$ the higher is the
plateau.

The viscosity data on microemulsion (Fig. \ref{2})
shows in fact such a plateau,
suggesting that the mechanism for the appearance of the plateau is
linked to the bond lifetime which in turn is related to the
cluster relaxation time.

\section{\bf Future Directions}
\label{future}

In conclusion, we have discussed the interplay between percolation
line and critical point in systems where thermal correlations play
an important role. The problem to define the droplets in spin models
is satisfactorily solved. However there are still some open problems. Above
$d = 4$ in the Ising model the definition of droplets presents some
difficulties, probably related to the upper critical dimension for
the percolation problem which is $6$. This type of difficulties does
not allow for a trivial extensions of the arguments used in the
random percolation problem, to explain the
hyperscaling breakdown. Another open problem is the characterization
of the thermal scaling exponent $1/\nu$, in terms of the fractal
dimension of some subset of the critical droplet, as occurs in the
random percolation problem.

In the last decade the KF,CK approach has been extended to
frustrated systems. Interestingly this approach has led to a new
frustrated percolation model, with unusual properties relevant to
spin glasses and other glassy systems \cite{frustrated,spin_glasses}.
However the precise
definition of clusters, which are able to characterize the
critical droplets for spin glasses, is still missing.

Although some advances have been obtained towards a
droplet definition in Lennard Jones systems \cite{campisator}, a general
definition for continuum models of fluids  still needs to be formulated.
\newpage

\vspace{2cm}

\vspace{2cm}

\vspace{2cm}

\vspace{2cm}

\vspace{2cm}

\newpage
\appendix
\section{\bf Random Cluster Model and Ising droplets}
\label{KF}
In 1969 Kasteleyn and Fortuin (KF) \cite{3} introduced a correlated
bond percolation model, called random cluster model, and showed that
the partition function of this percolation model was identical to
the partition function of $q-$state Potts model. They also showed that
the thermal quantities in the Potts model could be expressed in
terms of connectivity properties of the random cluster model. Much
later in 1980 Coniglio and Klein \cite{CK} independently have used a
different approach with the aim to define the proper droplets in the
Ising model. It was only later that it was realized that the two
approaches were related, although the meaning of the clusters in the
two approaches is different. We will discuss these two approaches
here, and show that their statistical properties are the same.

\subsection{\bf Random Cluster Model}

Let us consider an Ising system of spins
$S_{i}={\pm }1$ on a lattice with nearest-neighbour interactions
and, when needed, let us assume periodic boundary conditions in both
directions. All interactions have strength $J$ and the Hamiltonian
is
\begin{equation}
\mathcal{H}(\{S_{i}\})=-\sum_{\langle i,j \rangle
}J(S_{i}S_{j}-1)\,, \label{Ham}
\end{equation}
where $\{S_{i}\}$ represents a spin configuration and the sum is
over $nn$ spins. The main point in the KF approach is to replace the
original Ising Hamiltonian with an annealed diluted Hamiltonian
\begin{equation}
\mathcal{H}^{\prime }(\{S_{i}\})=-\sum_{<i,j>}J_{ij}^{\prime
}(S_{i}S_{j}-1)\, ,  \label{Hamd}
\end{equation}
where
\begin{equation}
J_{ij}^{\prime }=\left\{
\begin{array}{ll}
J^{\prime } & \mbox{with probability}~p \\
~ & ~ \\
0 & \mbox{with probability}~(1-p)\ .
\end{array}
\right.
\end{equation}

The parameter $p$ is chosen such that the Boltzmann factor
associated to an Ising configuration of the original model coincides
with the weight associated to a spin configuration of the diluted
Ising model
\begin{equation}
e^{-\beta \mathcal{H}(\{S_i\})} \equiv \prod_{<i,j>} e^{\beta
J(S_iS_j-1)}= \prod_{<i,j>}\left( pe^{\beta
J^{\prime}(S_iS_j-1)}+(1-p)\right) \, , \label{W_Ising}
\end{equation}
where $\beta=1/k_BT$, $k_B$ is the Boltzmann constant and $T$ is the
temperature. In order to satisfy (\ref{W_Ising}) we must have
\begin{equation}
e^{\beta J(S_iS_j-1)}=pe^{\beta J^{\prime}(S_iS_j-1)}+(1-p) \ .
\label{condition}
\end{equation}
We take now the limit $J^{\prime}\mapsto\infty$. In such a case
$e^{\beta J^{\prime}(S_iS_j-1)}$ equals the Kronecker delta
$\delta_{S_iS_j}$ and from (\ref{condition}) $p$ is given by
\begin{equation}
p=1-e^{-2\beta J} .  \label{A_p}
\end{equation}
From (\ref{W_Ising}), by performing the products we can write
\begin{equation}
e^{-\beta \mathcal{H}(\{S_i\})} = \sum_C W_{KF}(\{S_i\},C) \, ,
\label{Wkf}
\end{equation}
where
\begin{equation}
W_{KF}(\{S_i\},C)=p^{|C|}(1-p)^{|A|}\prod_{<i,j>\in
C}\delta_{S_iS_j}\, . \label{kfweight}
\end{equation}
Here $C$ is a configuration of interactions
where $|C|$ is the number of interactions of
strength $J^{\prime}=\infty$ and $|A|$ the number of interactions of
strength $0$. $|C|+|A|=|E|$, where $|E|$ is the
total number of edges in the lattice.

$W_{KF}(\{S_i\},C)$ is the statistical weight associated a) to a
spin configuration $\{S_i\}$ and b) to a set of interactions in the
diluted model where $|C|$ edges have
$\infty$ strength interactions, while all the other edges have $0$ strength
interactions. The Kronecker delta indicates that two spins connected
by an $\infty$ strength interaction must be in the same state.
Therefore the configuration $C$ can be decomposed in clusters of
parallel spins connected by infinite strength interactions.

Finally the partition function of the Ising model $Z$ is obtained by
summing the Boltzmann factor (\ref{Wkf}) over all the spin
configurations. Since each cluster in the configuration $C$ gives a
contribution of $2$, we obtain:
\begin{equation}
Z = \sum_C p^{|C|}(1-p)^{|A|} 2^{N_C} \, ,  \label{zkf}
\end{equation}
where $N_C$ is the number of clusters in the configuration $C$.

In conclusion, in the KF formalism the partition function
(\ref{zkf}) is equivalent to the partition function of a correlated
bond percolation model \cite{3, hu} where the weight of each bond
configuration $C$ is given by
\begin{equation}
W(C) = \sum_{\{ S_i\}}W_{KF}(\{S_i\},C) = p^{|C|}(1-p)^{|A|} 2^{N_C}
\label{wkf1}
\end{equation}
which coincides with the weight of the random percolation except for
the extra factor $2^{N_C}$. Clearly all percolation quantities in
this correlated bond model weighted according to Eq. (\ref{wkf1})
coincide with the corresponding percolation quantities of the KF
clusters made of parallel spins connected by $\infty$ strength
interaction, whose statistical weight is given by (\ref{kfweight}).
Moreover using (\ref{kfweight}) and (\ref{Wkf})
Kasteleyn and Fortuin have proved that \cite{3}
\begin{equation}
| \langle S_i \rangle | = \langle \gamma_i^{\infty}\rangle_{W}
\end{equation}
and
\begin{equation}
\langle S_iS_j\rangle=\langle\gamma_{ij}\rangle_{W} \, ,
\end{equation}
where $\langle ... \rangle$ is the Boltzmann average and $\langle
...\rangle _{W}$ is the average over bond configurations in the bond
correlated percolation with weights given by (\ref{wkf1}). Here $
\gamma_i^{\infty}(C)$ is equal to 1 if the spin at site $i$ belongs
to the spanning cluster, $0$ otherwise; $\gamma_{ij}(C)$ is equal to
1 if the spins at sites $i$ and $j$ belong to the same cluster, $0$
otherwise.

\subsection{\bf Connection between the Ising droplets and the
Random Cluster Model}

In the approach followed by Coniglio and Klein \cite{CK}, given a
configuration of spins, one introduces at random connecting bonds
between $nn$ parallel spins with probability $p_b$, antiparallel
spins are not connected with probability $1$. Clusters are
defined as maximal sets of parallel spins connected by bonds. The
bonds here are fictitious, they are introduced only to define the
clusters and do not modify the interaction energy as in the FK
approach. For a given realization of bonds we distinguish the
subsets $C$ and $B$ of $nn$ parallel spins respectively connected
and not connected by bonds and the subset $D$ of $nn$
antiparallel spins. The union of $C$, $B$ and $D$ coincides with
the total set of $nn$ pair of spins $E$. The statistical
weight of a configuration of spins and bonds is \cite{coniglio90,
conigliodiliberto}
\begin{equation}
W_{CK}(\{S_i\},C) = p_b^{\mid C\mid}(1-p_b)^{\mid
B\mid}e^{-\beta\mathcal{H} (\{S_i\})}  \, , \label{Conweight}
\end{equation}
where $|C|$ and $|B|$ are the number of $nn$ pairs of parallel
spins respectively in the subset $C$ and $B$ not connected by
bonds.

For a given spin configuration, using Newton binomial rule, we have the
following sum rule
\begin{equation}
\sum_{C} p_b^{\mid C\mid}(1-p_b)^{\mid B\mid}=1 .  \label{0.25}
\end{equation}
From Eq. (\ref{0.25}) follows that the Ising partition function,
$Z$, may be obtained by summing (\ref{Conweight}) over all bond
configurations and then over all spin configurations.
\begin{equation}
Z=\sum_{\{S_i\}}\sum_C W_{CK}(\{S_i\},C) = \sum_{\{S_i\}} e^{-\beta\mathcal{H%
}(\{S_i\})} \ .  \label{noname14}
\end{equation}

The partition function of course does not depend on the value of
$p_b$ which controls the bond density. By tuning $p_b$ instead it
is possible to tune the size of the clusters. For example by
taking $p_b=1$ the clusters would coincide with nearest neighbour
parallel spins, while for $p_b=0$ the clusters are reduced to
single spins. By choosing the droplet bond probability 
$p_b=1-e^{-2\beta J}\equiv p$  and observing that $e^{-\beta\mathcal{H}%
(\{S_i\})}=e^{-2\beta J|D|}$, where $|D|$ is the number of
antiparallel pairs of spins, the weight (\ref{Conweight})
simplifies and becomes:
\begin{equation}
W_{CK}(\{S_i\},C) = p^{\mid C\mid}(1-p)^{\mid A\mid}\, ,  \label{Conweight1}
\end{equation}
where $|A|=|B|+|D|=|E|-|C|$. 

From (\ref{Conweight1}) we can calculate the weight $W(C)$ that a
given configuration of connecting bonds $C$ between $nn$ parallel
spins occurs. This configuration $C$ can occur in many
spin configurations. So we have to sum over all spin
configurations compatible with the bond configuration $C$, namely
\begin{equation}
W(C)=\sum_{\{S_{i}\}}W_{CK}(\{S_{i}\},C)\prod_{<i,j>\in C}\delta
_{S_{i}S_{j}} \, , \label{ckkfweight1}
\end{equation}
where, due to the product of the Kronecker delta, the sum is over all spin
configurations compatible with the bond configuration $C$. From (\ref
{Conweight1}) and (\ref{ckkfweight1}) we have
\begin{equation}
W(C)=\sum_{\{S_{i}\}}p^{\mid C\mid }(1-p)^{\mid A\mid
}\prod_{<i,j>\in C}\delta
_{S_{i}S_{j}}=p^{|C|}(1-p)^{|A|}2^{N_{C}} \, . \label{ckkfweight2}
\end{equation}

Consequently in (\ref{noname14}) by taking first the sum over all spins 
compatible with the configuration $C$, the partition function $Z$ can be 
written as in the KF formalism (\ref{zkf}).
\begin{equation}
Z = \sum_C p^{|C|}(1-p)^{|A|} 2^{N_C} .  \label{zkf2}
\end{equation}

In spite of the strong analogies the CK clusters and the KF clusters
have a different meaning. In the CK formalism the clusters are
defined directly in a given configuration of the Ising model as
parallel spin connected by fictitious bonds, while in the KF
formalism clusters are defined in the equivalent random cluster
model. However, due to the equality of the weights
(\ref{Conweight1}) and (\ref{kfweight}) the statistical properties
of both clusters are identical \cite{conigliodiliberto} and due to
the relations between (\ref {kfweight}) and (\ref{wkf1}) both
coincide with those of the correlated bond percolation whose weight
is given by (\ref{wkf1}). More precisely, any percolation quantity
$g(C)$ which depends only on the bond configuration has the same
average
\begin{equation}
\langle g(C)\rangle _{KF}=\langle g(C)\rangle _{CK}=\langle
g(C)\rangle _{W} \, ,\label{averages}
\end{equation}
where $\langle ...\rangle _{KF}$, $\langle ...\rangle _{CK}$ are the
average over spin and bond configurations with weights given by
(\ref{kfweight}) and (\ref{Conweight1}) respectively and $\langle
...\rangle _{W}$ is the average over bond configurations in the bond
correlated percolation with weights given by (\ref{wkf1}). In view
of (\ref{averages}) it follows \cite{conigliodiliberto}
\begin{equation}
| \langle S_i \rangle | = \langle \gamma_i^{\infty}\rangle_{CK}
\label{Sgammainfinity}
\end{equation}
and
\begin{equation}
\langle S_iS_j\rangle=\langle\gamma_{ij}\rangle_{CK} \, .
\label{SSgamma}
\end{equation}

We end this section noting that in order to generate an equilibrium
CK droplet configuration in a computer simulation, it is enough to
equilibrate a spin configuration of the Ising model and then
introduce at random fictitious bonds between parallel spins with a
probability given by (\ref{A_p}).

\newpage
\begin{itemize}
\item[]{\bf Primary Literature}
\end{itemize}

\begin{itemize}
\item[]{\bf Books and Reviews}
\item[] Grimmett GR (1989) Percolation. Springer, Berlin
\item[] Sahimi M (1994) Application of Percolation Theory. Taylor and Francis 
\end{itemize}
\end{document}